\documentclass{pasa}%

\title[Variability in Ultra-luminous X-ray Sources]{Variability in Ultra-luminous X-ray Sources}
\author[Webb et al.]{N. A. Webb$^{1,2}$, D. Cseh$^3$ \and F. Kirsten$^4$ \\
\affil{$^1$Universit\'e de Toulouse; UPS-OMP; IRAP;  Toulouse, France}
\affil{$^2$CNRS; IRAP; 9 av. Colonel Roche, BP 44346, F-31028 Toulouse cedex 4, France}
\affil{$^3$Department of Astrophysics/IMAPP, Radboud University Nijmegen, P.O. Box 9010, 6500 GL Nijmegen, The Netherlands}
\affil{$^4$Max Planck Institut f\"ur Radioastronomie (MPIfR), Auf dem H\"ugel 69, D-53121, Bonn, Germany}}
\jid{PASA}
\doi{10.1017/pas.\the\year.xxx}
\jyear{\the\year}

\begin{document}%
\begin{abstract}
Many upcoming surveys, particularly in the radio  and optical domains, are designed to probe either the temporal and/or the spatial variability of a range of astronomical objects. In the light of these high resolution surveys, we review the subject of ultra-luminous X-ray (ULX) sources, which are thought to be accreting black holes for the most part.  We also discuss the sub-class of ULXs known as the hyper-luminous X-ray sources, which may be accreting intermediate mass black holes. We focus on some of the open questions that will be addressed with the new facilities, such as the mass of the black hole in ULXs, their temporal variability and the nature of the state changes, their surrounding nebulae and the nature of the region in which ULXs reside.
\end{abstract}
\begin{keywords}
stars: black holes -- accretion, accretion discs -- X-rays: binaries  
\end{keywords}
\maketitle%
\section{INTRODUCTION }
\label{sec:intro}
Ultra-luminous X-ray sources (ULXs) are X-ray
sources located outside the nucleus of their host galaxy, with luminosities that exceed the Eddington luminosity ($\rm L_{Edd}$) of a stellar mass black hole (BH), where $\rm L_{Edd}$ = 1.3 $\times$ 10$^{38}$(M/M$_\odot$) erg
s$^{-1}$.  In the literature, an off-nuclear source is often considered to be a ULX if L $>$ 1.3 $\times$ 10$^{39}$ erg s$^{-1}$.  These enigmatic sources were first discovered with the {\em Einstein} observatory between 1978 and 1981 \cite{fabi89} and their nature was unclear. Fabianno \shortcite{fabi89} did state that if these sources were indeed accreting compact objects, they would indicate massive (stellar mass) black holes.  The high luminosities, in conjunction with observations that showed
various ULXs in different spectral states, along with the fact that some ULXs appeared to
show spectral state transitions \cite{kubo01}, supported the idea that most
ULXs are accreting black holes.  

The masses of the ULX black holes are still somewhat under debate.  If we
assume that the maximum radiative luminosity from the
spherical accretion of matter is in indeed at the Eddington limit, the high luminosities observed from ULXs would imply that these sources are the previously unobserved and long sought after intermediate mass black holes (IMBH, $\sim$10$^2$ to $\sim$10$^5$ M$_{\odot}$).  Such sources are believed to exist as it is thought that mergers of IMBHs could be at the origin of supermassive black holes e.g. Madau \& Rees \shortcite{mada01}, or that super-Eddington accretion onto smaller mass
BHs could allow them to grow into supermassive black holes, via the IMBH state \cite{kawa04}. How IMBHs form and where they are located are two of a number of open questions concerning IMBH \cite{mill04}. They could be formed through the accretion of gas or smaller stars and therefore reside in the centres of old dense stellar clusters (globular clusters) or they may be formed directly in young star forming regions e.g. Miller \& Colbert \shortcite{mill04}. IMBHs are therefore the missing BH link.  Further, they are interesting for a diverse range of fundamental physics and astrophysical subjects. It is has been proposed that they may inject essential energy into globular clusters to slow down the eventual core collapse, e.g. Hut et al. \shortcite{hut92}, that dark matter may be relatively simple to detect around IMBH, e.g.  Fornasa \& Bertone \shortcite{forn08}, that IMBHs may have played a role in the cosmological ionisation \cite{mada04}, or that they may be strong sources of gravitational waves if the IMBH has a compact object companion in an elliptical orbit around it \cite{mill04} and that gas accretion onto IMBHs should have made a significant contribution to the UV background \cite{kawa03}.

However, not all ULXs can host black holes of
such a high mass.  For instance, King \shortcite{king04} showed that the
population of ULXs in the cartwheel galaxy is difficult to explain if
they are all IMBHs, as much of  the star forming
mass would then end up as IMBHs.
Alternatively, if the ULX emission is collimated, possibly through a
geometrically thick accretion disc \cite{king01} or via
relativistic boosting \cite{koer02}, the emission could simply appear
to exceed the Eddington limit. Indeed a number of Galactic black hole
binaries are seen to surpass such a limit, i.e. V4641 Sgr
\cite{hjel00} and GRS 1915+105 \cite{done04}.  Luminosities up to $\sim$10$^{41}$
erg s$^{-1}$ can be plausibly explained through beaming effects \cite{pout07,free06} and/or hyper-accretion onto stellar mass BHs \cite{pout07,bege02}. Therefore, the majority of ULXs are thought to host BHs with masses close to those of stellar mass black holes \cite{robe07}. However, a rare class of ULXs -- the hyper-luminous
X-ray sources (HLX) -- emit X-rays at luminosities $>$10$^{41}$ erg s$^{-1}$  \cite{gao03}
and require increasingly complicated scenarios to explain them without
invoking the presence of an IMBH.

Few candidate HLX have been observed, but one excellent candidate is a 2XMM X-ray catalogue source \cite{wats09} 2XMM J011028.1460421,
discovered as an off-nuclear X-ray source seemingly associated with
the galaxy ESO 243-49, 95 Mpc (z = 0.0224) away
\cite{farr09}. The optical counterpart located by Soria et al. \shortcite{sori10} using the excellent X-ray position derived from Chandra data \cite{webb10}, shows an H$_\alpha$ emission line commensurate with the galaxy distance, thus confirming the association with ESO 243-49 \cite{wier10}. Taking the maximum X-ray luminosity of 1.1 $\times$ 10$^{42}$ erg s$^{-1}$
(0.2-10.0 keV, unabsorbed) and assuming that
this value exceeds the Eddington limit by a factor of 10,
implies a minimum mass of 500 M$_\odot$ \cite{farr09}. Modelling and Eddington
scaling of the X-ray data imply a mass of $\sim$10$^4$ M$_\odot$ \cite{gode12,davi11,serv11} and recent radio
observations yield an upper mass limit of 9 $\times$ 10$^4$
M$_\odot$, supposing Eddington scaling \cite{webb12}. This source goes under the name of ESO 243-49 HLX-1 or HLX-1 for
short.

\subsection{ULXs and their different states}

Many ULXs are seen to be variable, particularly in the X-ray domain, e.g. Kubota et al. \shortcite{kubo01} and Kajava \& Poutanen \shortcite{Kajava:2009sj}. In addition, the fact that most ULXs are accreting black holes means that they are frequently likened to Galactic black hole X-ray binaries (GBHB), albeit extreme versions. GBHBs show both long and short term X-ray variability.  On the long term, they exhibit well defined X-ray states: the high/soft (or
thermal), low/hard (or hard) and very high (or steep power law) states, see for instance Remillard \& McClintock \shortcite{MR06}.  Much work has gone into understanding if ULXs behave in a similar way to GBHBs \cite{glad09,Kajava:2009sj,robe12}.  Such studies indicate that the majority of ULXs show a more extreme version of the GBHB states, going beyond the high state and into super-Eddington accretion regimes, known as the ultra-luminous accretion state \cite{glad09b}, see also Sec.~\ref{sec:var_state}.  Observing these states, and 
transitions between them, could provide important information on how the Eddington limit can be broken, which may require new physics. Super Eddington accretion has been invoked to explain how the earliest supermassive black holes were formed e.g. Kajava et al \shortcite{Kajava:2009sj}.   

Some ULXs, however, have been shown to exhibit the canonical X-rays states observed from GBHBs, the first of these being HLX-1, see Sec.~\ref{sec:states}. This may indicate that ULXs can be divided into sub-classes, namely with sub and super-Eddington emission.  The implications of this are discussed further in  Sec.~\ref{sec:states}.

\subsection{ULX nebulae}

Some ULXs show optical emission-line nebulae, due to shock-ionised driven jets, outflows or disc winds and/or because of photo-ionisation from the X-ray and UV emission around the black hole. Some of these also show a radio counterpart. These nebulae can
be used as a calorimeter to infer the total intrinsic power of
the ULX, \cite{paku02,paku03}, and also to understand how outflows and photo-ionisation can play a role in the behaviour of the ULX and on its surrounding environment \cite{paku02,paku03,robe03,kaar04b}.

\subsection{Spatial and temporal variability with upcoming instrumentation}

With the wide range of new and forthcoming observatories and surveys over the next few years, there will be a huge amount of data available to study open questions surrounding ULXs. The new hard X-ray observatory, {\em Nuclear Spectroscopic Telescope Array} ($NuSTAR$), launched on 2012 June 13 operates in the band 3-79 keV  \cite{harr13}. Compared to other X-ray satellites such as {\em XMM-Newton} and {\em Chandra}, that only probe the spectrum up to $\sim$12 keV, {\em NuSTAR}'s extended X-ray spectral band allows us, for the first time, to probe ULX spectra above 10 keV with high signal to noise. This should allow us to constrain models describing the ultra-luminous state in the near future. It will also allow us to uncover new ULXs (e.g. Walton et al. 2013b), thanks to the large field of view (FOV, 10' at 10 keV, 50\% response), that were previously veiled by strong interstellar absorption, maybe revealing a new shrouded population of ULXs, for example.  

In the optical, future surveys such as the {\em Large Synoptic Survey Telescope} ($LSST$)\footnote{http://www.lsst.org/lsst/} will observe the sky using an 8.4 m telescope. The observations will be made using 6 bands (0.3-1.1 micron) and have  a 3.5$^\circ$ FOV. It is expected to be operating in 2022 and will regularly scan the sky, so it will be ideal for picking out new and monitoring known transient ULXs. This will help to identify ULXs under-going state transition, in order to understand how these processes relate to those in GBHBs, see Sec.~\ref{sec:var_state}. 

At longer wavelengths still, the new radio arrays that are coming on line in the near future will allow us to probe the same region of the sky repeatedly, to depths that equal the most sensitive radio telescopes currently in use.  {\em The Hunt for Dynamic and Explosive Radio Transients with Meerkat} ($Thunderkat$) project will devote 3000 h of time between 2016 and 2020 on the 64 dishes that comprise {\em Meerkat} (0.9-1.7 GHz)\footnote{http://www.ast.uct.ac.za/thunderkat/ThunderKAT.html}. This high sensitivity observatory will allow a 5 $\sigma$ detection of a faint radio source (14 $\mu$Jy) in just 8 hours. $Thunderkat$ will be devoted to following up transient phenomena as a compliment to the regular {\em Meerkat} program. Spectral changes along with spatial and temporal variability in the radio nebulae around ULXs will be probed and new radio ULXs will be discovered with such a program. Similarly, {\em VAST}, {\em An ASKAP Survey for Variables and Slow Transients}, \cite{murp13} will have a root mean square (RMS) continuum sensitivity of 47 $\mu$Jy beam$^{-1}$ in just 1 hr, thanks to its 36 {\em Australian Square Kilometre Array Pathfinder} ($ASKAP$) dishes. It will have a 30$^\circ$ FOV and will conduct broad and deep observations to achieve similar goals, starting as early as 2012.

Almost all of the above cited projects rely on detecting variability, on temporal or spatial scales, to identify new ULXs and eventually respond to the open questions cited above. Identifying more ULXs is essential to understand their demographics, especially those that are thought to contain IMBH, as how they form and evolve and whether they exist in globular clusters are still very much open questions, see Secs.~\ref{sec:intro} and \ref{sec:reside}. Further, if ULXs are black holes accreting from a high mass companion, many will evolve to form compact object binaries e.g. Mandel et al. \shortcite{mand08}.  Whilst it will be possible to detect such binaries using the electromagnetic spectrum, the new gravitational wave detectors {\em Advanced LIGO} and {\em Advanced  Virgo} should allow us to detect the inspiral of these binaries for which the BH has a  mass greater than 50 M$_\odot$ and up to 10000 M$_\odot$ \cite{mand08,gair11}. Many of the current studies aimed at identifying the numbers of such binaries use the estimated population of ULXs e.g. Mandel et al. \shortcite{mand08}. It is therefore through determining the population of ULXs that we should get a handle on the number of compact object binaries that we will detect with the new gravitational wave detectors.

In the rest of this review we will look at some specific aspects of ULX variability and how the new facilities coming on line will enable us to respond to questions surrounding ULXs.

\section{Variability due to state changes}
\label{sec:var_state}

X-ray spectral changes in conjunction with X-ray luminosity changes
have been observed in a number of ULXs, e.g. Kubota et
al. \shortcite{kubo01}. However, contrary to the GBHBs, the harder
state is often associated with the higher luminosity and the
softer state with the lower luminosity. This lead to the fourth
GBHB state being proposed, the {\em apparently standard regime}
\cite{kubo04} or {\em ultra-luminous state} \cite{glad09b}, when the BH
accretes at or above the Eddington limit. However, the physics of this state are not yet understood e.g. Walton et al. \shortcite{walt13}.  Variability of a factor of about 10 or more in X-rays can be seen during these state changes. Recently Soria et al. \shortcite{sori12b} observed a previously undetected (L $<$ 10$^{36}$ erg s$^{-1}$) object enter the ULX state (L $\sim$ 4$\times$10$^{39}$ erg s$^{-1}$), an X-ray flux change of more than three orders of magnitude. More recently, a new X-ray binary in M~31 was discovered \cite{henz12}, which peaks at L$_x$ $>$ 1 $\times$ 10$^{39}$ erg s$^{-1}$, qualifying it as a ULX. Its X-ray emission shows typical ULX states, but radio emission varying on minute timescales has been observed at the mJy level \cite{midd13}.  Another source also showing a transition to the ULX state, and with a variability of at least a factor 100, has been found in M~94 \cite{lin13}.  How and why this transition occurs is unclear, but 
 it may be that super-Eddington discs remain slim and accelerate a significant wind with a 'thick disc' geometry, unlike in the standard scenario where accretion discs become thick e.g. Dotan \& Shaviv \shortcite{dota11}.

It should be noted, however, that an X-ray
variability of a factor of a few can be seen with no
spectral state change, see e.g. Soria et al. \shortcite{sori09}.  The variability and the spectral changes are therefore poorly understood and further observations of known and new sources are essential to understand the behaviour of ULXs, especially in the ultra-luminous regime.

\begin{figure*}
\begin{center}
\includegraphics[width=20pc, height=42pc, angle=-90]{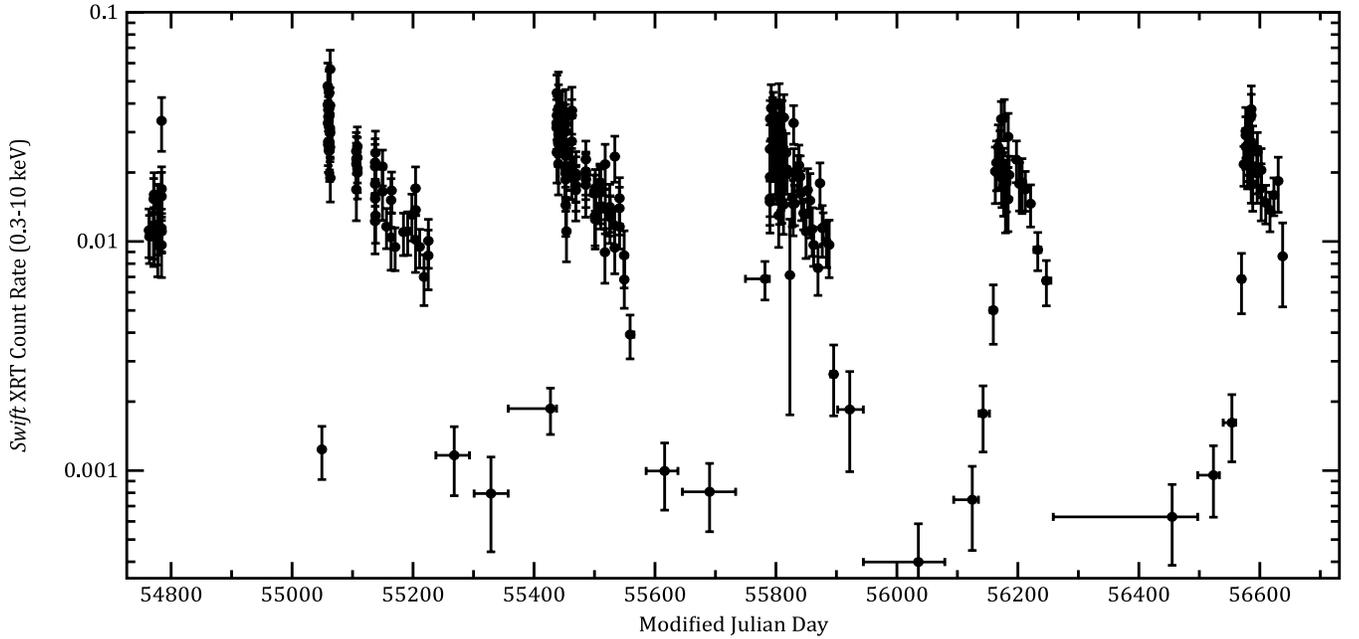}
\caption{The Swift X-ray lightcurve of ESO 243-49 HLX-1 from 2008 until the end of 2013}
 \label{fig:HLX1lc}
\end{center}
\end{figure*}

\subsection{Canonical black hole states in ULXs}
\label{sec:states}

The first ULX to show all three GBHB X-ray states, was HLX-1. The source has shown five fast rise and exponential decay (FRED) type X-ray outbursts, characterised by an increase in the count rate by a factor 40 \cite{gode12,gode09,laso11} and see Fig.~\ref{fig:HLX1lc}. From simple spectral fitting, the unabsorbed luminosities range from 1.9 $\times$ 10$^{40}$ to 1.3 $\times$ 10$^{42}$ erg s$^{-1}$. At high luminosities, the X-ray spectrum shows a thermal state dominated by a disc component with temperatures $\le$0.26 keV. At low luminosities the
spectrum is dominated by a hard power law with 1.4 $\le \Gamma \le$ 2.1, consistent with
a hard state. The source has also been observed with a power law spectrum with $\Gamma$=3.3$\pm$0.2, consistent with the steep power law state \cite{farr09}. When HLX-1 is in the high/soft state, the luminosity of the disc component appears to
scale with the inner disc temperature to the power of four, which supports the presence of an optically
thick, geometrically thin accretion disc.  The observed spectral changes and long-term variability are not consistent with variations in the beaming of the emission, nor are they consistent with the source being
in a super-Eddington accretion state \cite{serv11}. Indeed, through modelling the X-ray spectra with the Kawaguchi \shortcite{kawa03} disc model and assuming a non-spinning black hole and an inclination of 0$^\circ$, the accretion flow was determined to be in the sub-Eddington regime for all observations \cite{gode12}, radiating close to the Eddington limit at the peak of the outburst. The disc radiation efficiency was shown to be
 $\eta$ = 0.11 $\pm$ 0.03 \cite{gode12}. 

In addition, radio flares have been detected from HLX-1 during the low/hard to high/soft state transitions in 2010 and 2011 \cite{webb12}, in a similar way to GBHBs, which are known to  regularly emit radio flares
around this transition,
e.g. \cite{fend09,corb04}. These jets are associated with 
ejection events, where, for example, the jet is expelled which can lead to radio flaring when the higher velocity ejecta may collide with the lower-velocity material produced by the steady jet or the local environment. This reinforces the similarity between HLX-1 and the GBHBs. Also, Webb et al. \shortcite{webb14} showed that the V-band optical emission increases from before the X-ray outburst and through the X-ray outburst by 2 magnitudes. This concurrent increase is similar to that observed in GBHBs and opens up the possibility that optical surveys will detect spectral state changes in ULXs. 

Other objects have been identified with X-ray luminosities $>$10$^{40}$ erg s$^{-1}$ that show hard power law spectra, e.g. Zolotukhin et al. \shortcite{zolo13}, and some with significant RMS variability \cite{sutt12}, typical of the low/hard state. These objects are not formally HLXs. However, if they are in the low/hard state, which is thought to occur at $\sim$1-2\% Eddington luminosity \cite{macc03}, and then they switch to the high state (i.e. close to the Eddington luminosity) they will have luminosities of the order of $\sim$10$^{42}$ erg s$^{-1}$ and so become HLXs. Such high luminosities indicate that they may contain IMBHs, see Sec.~\ref{sec:intro}. However, so far, no such state change has been observed in these objects and further observations will be required to confirm the IMBH nature.  If confirmed, this would provide an observational constraint for identifying ULXs containing IMBH.

\subsection{The Ultra-luminous state}
\label{sec:ulx_state}

Recent work with {\em NuSTAR} has focused on
understanding the underlying mechanism of the {\em ultra-luminous
  state}. {\em NuSTAR} observations of the two ULXs NGC 1313
X-1 and X-2, show a clear cutoff of the X-1's spectrum above 10 keV,
which was only marginally detectable with previous X-ray
observations. This cutoff rules out the interpretation of X-1 as a BH
in a standard low/hard state, or in a reflection-dominated state. The
cutoff is also found to differ from that which is predicted by a
single-temperature Comptonisation model \cite{bach13}. This is also the case for the new ULX, Circinus ULX5, discovered with {\em NuSTAR} \cite{walt13b}. A spectral transition in NGC 1313 X-2, from a state
with high luminosity and strong fractional variability that increases with energy to a lower-luminosity state with no detectable variability, was explained by a transition from a super-Eddington to the sub-Eddington regime \cite{bach13}. This type of transition has been observed in other objects, e.g. Burke et al. \shortcite{burk13}.  Sutton et al. \shortcite{sutt13} showed that high levels of fractional variability are seen in ULXs with soft ultra-luminous spectra as well, likewise for a couple of the broadened disc sources. Furthermore, they observe variability that is strongest at high energies. They propose that such properties are consistent with models of super-Eddington emission, where a massive, radiatively driven wind forms a funnel-like geometry around the central regions of the accretion flow. Future observations with {\em NuSTAR} should help to confirm or refute such a hypothesis.

\section{Where do ULXs reside?}
\label{sec:reside}

ULXs are found preferentially in star forming regions e.g. Grimm et al. \shortcite{grim03}, but there is also evidence to show that ULXs form in low metallicity regions e.g. Pakull \& Mirioni \shortcite{paku02}, Mapelli et al. \shortcite{mape09} and Prestwich et al. \shortcite{pres13}. Indeed two of the recently discovered ULXs, see Sec.~\ref{sec:var_state}, appear to have old stellar counterparts \cite{sori12b,midd13}, associating them with old, low metallicity stellar populations.  

If ULXs are simply BHs accreting from intermediate or high mass donor stars with the X-ray emission beamed towards us, see Sec.~\ref{sec:intro}, Irwin et al. \shortcite{irwi04} argued that ULXs should not be found in old stellar populations.  Irwin et al. \shortcite{irwi04} showed that there are indeed very few ULXs associated with early-type galaxies.  However, Maccarone et al. \shortcite{macc07} discovered a ULX in a globular cluster (typically an old population) in the giant elliptical galaxy NGC 4472, thanks to its strong X-ray variability. Since then, at least five more ULXs have been proposed in extra-galactic globular clusters \cite{macc11,irwi10,shih10,bras12,robe12}. Whilst this may seem to support the idea that ULXs are found in metal poor regions, Maccarone et al. \shortcite{macc11} found that the five globular cluster ULXs known at the time, were found in metal rich clusters.  

No ULXs have so far been detected in Galactic globular clusters, but Strader et al. \shortcite{stra12b} did discover the first stellar mass black holes using radio observations. These black holes may well be slightly more massive than other Galactic stellar mass black holes, but further observations will be required to confirm this possibility \cite{stra12b}.  Why no ULXs exist in Galactic globular clusters is so far unexplained, but it may be related to the metallicity \cite{mape13} or simply because the Galactic globular cluster system is small and thus no such objects may be expected.

With regards to the HLX population, Farrell et al. \shortcite{farr12} showed through modelling the X-ray and near UV to near IR light that HLX-1 may reside in an old stellar population, akin to a globular cluster, although the preferred solution is a young stellar population \cite{farr13}.  None of the other proposed HLXs have been reliably associated with a host stellar population.  

Black hole mass scaling relations e.g. L\"utzgendorf et al. \shortcite{lutz13} and
stellar velocity dispersion measurements in globular clusters e.g. Anderson \& van der Marel \shortcite{ande10}, suggest the existence of IMBHs near
or exactly at the potential centre of globular clusters. If these objects are accreting material, they should appear as ULXs/HLXs (depending on the accretion rate). Searches for such objects in Galactic globular clusters using X-ray observations have not revealed such a source e.g. Pooley \& Rappaport \shortcite{pool06} and Pooley et al. \shortcite{pool03}.  If, however, the BH was accreting from gas in the globular cluster, the emission of photons might be
inefficient in the X-ray regime. BHs would then appear as compact
radio sources co-moving with the globular cluster.  Searches for sufficient quantities of gas in Galactic globular clusters have been fruitful. Despite the low gas density,
Freire et al. \shortcite{frei01} detected a non-negligible amount of ionized gas in 47 Tuc through pulsar timing observations. M~15 also shows evidence for dust \cite{boye06} and only a small fraction ($\sim$10\%, Maccarone \shortcite{macc04}) is required to be
accreted by a putative IMBH for it to become visible in the
X-ray/radio domain.  

Radio flares that can accompany state changes  should also be
detectable in long term radio observations. To date, neither a
continuous nor a flaring radio source has been detected down to
the sensitivity limits of current instruments e.g. Maccarone \& Servillat \shortcite{macc08}, Kirsten \& Vlemmings \shortcite{kirs12} and Strader et al. \shortcite{stra12}. This might imply IMBHs
simply do not exist in globular clusters or that gas produced by AGB stars in globular clusters does not survive
long enough to be accreted by the central source, hence prohibiting the
formation of an accretion disc and relativistic jets \cite{boye06}.

\section{ULXs in future surveys}

The X-ray variability  allows
new ULXs to be identified using X-ray catalogues such as the {\em
  2XMM} catalogue of serendipitously detected {\em XMM-Newton} sources
\cite{wats09}, e.g. Lin, Webb \& Barret \shortcite{lin12}. The recently released {\em
  3XMM} catalogue includes 372728 unique X-ray sources distributed across 650 square degrees ($\sim$1.5\%) of sky, almost 50\% more sources than in the previous version. Over 123800 of these sources are provided with both spectra and lightcurves. {\em 3XMM} also has the added value that it includes complete uniform reprocessing of all of the data, that exploits the most recent processing and calibration improvements. This is then a fantastic resource for searching for new ULXs.

However, it is likely that within the next ten years the existing sensitive X-ray satellites (e.g. {\em Swift, XMM-Newton, Chandra}) will no longer be in operation. In the 2020's, it is possible that there will be no new big X-ray telescope to take their place before the launch of {\em Athena+} at the end of the 2020's  \cite{nand13}. X-ray telescopes have historically been used to detect and identify new ULXs, often through the variability due to state changes, see Sec.~\ref{sec:var_state}. Without X-ray telescopes surveying the sky, the optical telescopes will take over this role, searching for the optical variability associated with state change and the intrinsic optical variability, both described in Sec.~\ref{sec:var_state}.

Identifying new ULXs through optical observations should eventually lead to the discovery of optical counterparts that are sufficiently bright that short spectroscopic exposures will achieve optical spectra with high enough signal to noise to spectrally identify the companion and to carry out a radial velocity studies, to measure the mass function. This will allow an accurate mass determination of the black hole to finally settle the debate on the mass of the back holes in these objects.   Previous attempts to detect a spectrum from the companion have revealed blue, almost featureless spectra \cite{zamp04,kaar09,robe11,gris11,cseh11,gris12} which are probably dominated by the accretion disc and in some cases show evidence of the surrounding nebula.   Some attempts to use disc lines \cite{cseh13,motc11} revealed no mass constraints. This may be because of inadequate time sampling of the radial velocity curve, too low
spectral resolution resulting in blurred narrow and
broad components, contrasting viewing geometries of the system,
or that the He II line is not emitted from the disc \cite{cseh13}.  M101 ULX-1 has however been shown to have a number of broad emission lines (FWHM$>$20 \AA), including He II 4686 \AA, HeI 5876 \AA\ and He I 6679 \AA, He I 4471 \AA, He I 4922 \AA\ and He II 5411 \AA\ lines and NIII 4640 \AA\ lines. The 
He I 5876/He II 5411 \AA\ equivalent width ratio suggests a Wolf-Rayet star. These lines have allowed Liu et al. \shortcite{liu13} to determine the mass of the Wolf-Rayet star (17.5 M$_\odot$) and the radial velocity, which reveals a black hole mass constraint of $>$5 M$_\odot$, but more likely in the range 20-30 M$_\odot$. 

The radio emission variability detected from HLX-1 (see Sec.~\ref{sec:states}) and the M~31 X-ray binary/ULX (see Sec.~\ref{sec:var_state}) opens up the possibility of discovering new, nearby, low-luminosity ULXs with the upcoming radio surveys.  The new radio surveys have very low detection limits (see Sec.~\ref{sec:intro}). Given that all the firm radio ULX counterparts have a total flux density below $\sim$1 mJy and sizes below 2-15'' at Mpc distances \cite{cseh12,webb12}, the forthcoming facilities should break the silence in systematic radio surveys looking for ULXs. 

Some current surveys, e.g. Legacy e-MERLIN 
Multi-band Imaging of Nearby Galaxies Survey (LeMMINGs), also search for extended radio emission from ULXs. However, lower resolution facilities 
may be able to measure flux variations of some of the already known or newly discovered 
radio bubbles around ULXs. Such variability is expected on the grounds of recent discoveries \cite{mezc13,cseh13b}, where a working jet is embedded in the radio bubble,
heavily perturbing it.

However, detecting distant (e.g. $>$100 Mpc) ULXs could still be challenging even with the new surveys. This is mainly due to the faint expected flux density of these sources and due to
the natural confusion limits; see e.g. the SKA \cite{jack04}. So, variability from future detections will have to trigger follow-up, high-resolution radio observations with VLBI.

With regards to spatial variability, measuring proper motions of ULXs \cite{kirs12} may be a future possibility
to independently infer the masses of these objects. Several ULXs are seen to be displaced from nearby star clusters \cite{kaar04} and they may have received natal 
velocity kicks that might be up to hundreds of km s$^{-1}$ for a 10 M$_{\odot}$ black hole \cite{mape13,repe12}. To detect such displacements would not just require better than milli-arcsecond resolution, but also perhaps, the detection of self-absorbed compact jets that are typically associated with hard, inefficient accretion states. This might be challenging in the light of the {\em NuStar} results, where it was shown that some ULXs are not in a canonical low/hard state, even though the X-ray spectrum is hard (see Sec.~\ref{sec:ulx_state}). None the less, recent VLBI observations of the ULX N5457-X9 in the low/hard state have revealed radio emission which is thought to be coming from compact jets \cite{mezc13b}, indicating that this should be possible, at least for some sources.




\end{document}